\documentclass[conference, 10pt, letterpaper]{IEEEtran}
\IEEEoverridecommandlockouts
\usepackage{amsfonts,cite,graphicx}

\usepackage[cmex10]{amsmath}
\newlength{\plotwidth}
\newtheorem{theorem}{Theorem}

\newtheorem{corollary}{Corollary}

\DeclareMathOperator*{\argmin}{arg\,min}

\begin{document}
\title{Energy Optimization across Training and Data for Multiuser Minimum Sum-MSE Linear Precoding}
\author{
\IEEEauthorblockN{Adam~J.~Tenenbaum and Raviraj~S.~Adve}
\IEEEauthorblockA{
Dept. of Electrical and Computer Engineering, University of Toronto\\
10 King's College Road, Toronto, Ontario, M5S 3G4, Canada\\
Email: \texttt{\{adam,rsadve\}@comm.utoronto.ca}}
\thanks{This work has been submitted to the IEEE for possible publication. Copyright may be transferred without notice, after which this version may no longer be accessible.}
}
\maketitle

\begin{abstract}
This paper considers minimum sum mean-squared error (sum-MSE) linear
transceiver designs in multiuser downlink systems with imperfect
channel state information. Specifically, we derive the optimal energy
allocations for training and data phases for such a system. Under MMSE
estimation of uncorrelated Rayleigh block fading channels with equal
average powers, we prove the separability of the energy allocation and
transceiver design optimization problems. A closed-form optimum energy
allocation is derived and applied to existing transceiver designs.
Analysis and simulation results demonstrate the improvements that can
be realized with the proposed design.
\end{abstract}

\section{Introduction}
Transceiver designs that minimize the sum of mean squared errors (sum-MSE) under a sum power constraint in the multiuser downlink with full channel state information (CSI) at the base station are well researched~\cite{SSJB05,KTA06,MJHU06,SSB07}.  In these papers, an uplink-downlink duality is used to transform a non-convex downlink problem into an equivalent convex virtual uplink problem.  Recent studies~\cite{DingPhD,SD08,DB09} have extended these original papers to the case of imperfect CSI, deriving an MSE duality in the presence of channel estimation errors and providing robust transceiver designs.

In order to design precoders, the base station must obtain estimates of the channel coefficients.  If channel reciprocity holds (i.e. the uplink and downlink channels are statistically identical), these estimates can be provided by training in the uplink (e.g., using uplink sounding, as in the WiMAX standard~\cite{80216e}).  However, in frequency division duplex systems (and in some broadband time division duplex systems~\cite{H08}), channel reciprocity does not apply.  In this case, channel estimation must be performed in the downlink and communicated back to the base station using an uplink feedback mechanism.  In this paper, we consider imperfect CSI estimation at the mobile receivers, but assume that the imperfect estimates are also available at the base station (via an error-free and delay-free feedback mechanism)\footnote{In this regard, this work complements~\cite{TAY08}, where we consider perfect receiver CSI estimates and a feedback mechanism incorporating prediction, error, and delay.}.

The algorithms designed in~\cite{DingPhD,SD08,DB09} for minimization of the sum-MSE under a sum-power constraint presume that fixed channel estimation error variances $\sigma_k^2$ are provided by a predetermined estimation mechanism.  In this paper, we address the problem of jointly designing a training sequence for MMSE CSI estimation and designing linear transceivers for minimum sum-MSE communication.  We consider the optimum allocation of limited available energy between the training and data communication phases for each quasi-static communication block. 

In Section~\ref{section:smbg}, we describe the channel model under consideration and review the design of training sequences for MMSE channel estimation.  We then present the linear precoding system model and provide an overview of the design of minimum sum-MSE linear precoders with imperfect CSI and fixed transmit power.  In Section~\ref{section:jointopt}, we formulate the joint design problem for energy allocation and precoder design.  We present a closed-form solution for the optimum training energy, and apply the result to existing precoder design techniques.  Performance and behaviour of the proposed approach are illustrated in Section~\ref{section:simresults}, and we draw conclusions in Section~\ref{section:conclusions}.  Appendix~\ref{section:mmseestimation} derives the MMSE channel estimation error variance and the calculations of our main proof are presented in Appendix~\ref{section:optTrainingEnergy}.

\emph{Notation}: We use the following conventions: italics represent scalars, lower case boldface type is used for vectors, and upper case boldface represents matrices, (e.g., $x,\mathbf{x},\mathbf{X}$, respectively).  Entries in
vectors and matrices are denoted as $\left[\mathbf{x}\right]_i$ and
$\left[\mathbf{X}\right]_{i,j}$. The superscripts $^T$ and $^H$ denote the
transpose and Hermitian operators. $\mathbb{E}[\cdot]$ represents
the statistical expectation operator while $\mathbf{I}_N$ is the $N\times N$
identity matrix. $\left\| \mathbf{x} \right\|_1$ and $\left\| \mathbf{x}\right\|_2$ denote the 1-norm (sum of entries) and Euclidean norm.
$\mathrm{diag}(\mathbf{x})$ represents the diagonal matrix formed using the
entries in vector $\mathbf{x}$, and
$\mathrm{diag}\left[\mathbf{X}_1,\ldots,\mathbf{X}_k \right]$ is the block
diagonal concatenation of matrices $\mathbf{X}_1,\ldots,\mathbf{X}_k$.  The $\mathrm{vec}(\mathbf{X})$ operator stacks the columns of the matrix $\mathbf{X}$ in a single vector.  $\mathcal{CN}(\mathbf{m},\mathbf{R})$ denotes the complex multivariate Gaussian probability distribution with mean $\mathbf{m}$ and covariance matrix $\mathbf{R}$.

\section{System Model and Background\label{section:smbg}}
\subsection{Channel Model}
\begin{figure}
\centering
\includegraphics[width=\plotwidth]{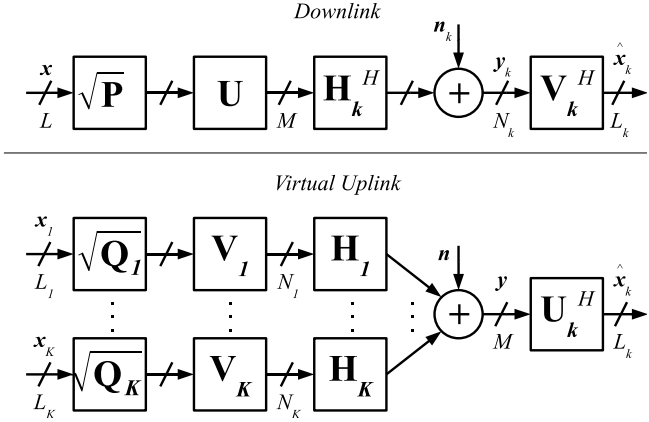}
\caption{Data processing for user $k$ in downlink and virtual uplink.}\label{fig:systemmodel}
\end{figure}

In the linear precoding system illustrated in Fig.~\ref{fig:systemmodel}, a base station with $M$ antennas transmits to $K$ decentralized mobile users with $N_k$ antennas each over flat wireless channels.
The channel between the transmitter and user $k$ is represented by the $N_{k}\times M$ matrix $\mathbf{H}_{k}^{H}$, and the overall $N\times M$ channel matrix is $ \mathbf{H}^H$, with
$\mathbf{H}=\left[\mathbf{H}_{1},\dots,\mathbf{H}_{K} \right]$, and where $N=\sum_k N_k$ is the total number of receive antennas in the system.  We assume that all channel coefficients are i.i.d. and drawn from a  zero-mean complex Gaussian distribution with variance $\sigma_H^2$; that is, $\mathrm{vec}(\mathbf{H}) \sim \mathcal{CN}(\mathbf{0},\sigma_H^2\mathbf{I}_{MN})$.  We consider a quasi-static (block fading) channel model, where the channel coefficients are assumed to be fixed for a coherence interval of $n$ consecutive symbol periods.  The first $n_T$ transmissions in each block are training symbols which the mobile receivers use to estimate the downlink channel, $\hat{\mathbf{H}}_k^H$; these imperfect CSI estimates are assumed to be available at the base station via an error-free and delay-free feedback mechanism.  We consider the stochastic error model (as used in~\cite{DingPhD,SD08,DB09}) where the true channel is modelled as a sum of the estimated channel and an independent additive error term, $\mathbf{H}_k=\hat{\mathbf{H}}_k + \mathbf{E}_k$ with $\mathrm{vec}(\mathbf{E}_k) \sim \mathcal{CN}(\mathbf{0},\sigma_k^2\mathbf{I}_{MN_k})$, and $\mathbf{E}=\left[\mathbf{E}_{1},\dots,\mathbf{E}_{K} \right]$.

\subsection{MMSE Channel Estimation and Training}
Training sequence and estimator design can be simplified under the assumption of uncorrelated channel coefficients by considering training for vector channels from the $M$ transmit antennas to each individual receive antenna.  To simplify notation in this section, we consider training for a single vector channel $\mathbf{h}^H$.  Channel estimation is performed by transmitting a set of $n_T$ training signal vectors, $\mathbf{X}_T = \left[\mathbf{x}_{T,1}, \dots, \mathbf{x}_{T,n_T}\right]$, from the $M$ transmit antennas without precoding.  $n_T \ge M$ training symbol vectors must be sent to guarantee resolvability of the individual channel coefficients.  The received signal vector is $\mathbf{y}_T = \mathbf{h}^H\mathbf{X}_T + \mathbf{z}$, where $\mathbf{z} \sim \mathcal{CN}(\mathbf{0},\sigma_n^2\mathbf{I}_{n_T})$, and the MMSE channel estimate $\hat{\mathbf{h}}_{\mathrm{MMSE}}^H=\mathbf{y}_T\mathbf{A}_0$ is found using the linear MMSE estimator $\mathbf{A}_0=\left(\mathbf{X}_T^H\mathbf{X}_T + \frac{\sigma_n^2}{\sigma_H^2}\mathbf{I}_{n_T}\right)^{-1}\mathbf{X}_T^H$.  Under the sum energy constraint, $\mathrm{tr}\left[\mathbf{X}_T^H\mathbf{X}_T\right] \le E_T$, where $E_T$ is the energy allocated to training, and the assumption of independent channel coefficients, a sufficient condition for optimality of the training matrix is $\mathbf{X}_T\mathbf{X}_T^H=\frac{E_T}{M}\mathbf{I}_M$~\cite{BG06}; that is, we are free to select any training matrix with orthogonal rows.  When using the MMSE estimator, there is no benefit using any more than $n_T=M$ training symbols.  For algorithmic simplicity, we choose the set of training vectors $\mathbf{X}_T = \sqrt{\frac{E_T}{M}}\mathbf{I}_M$.  One may also choose $\mathbf{X}_T$ as the scaled size-$M$ DFT matrix, $\left[\mathbf{X}_T\right]_{m,n}=\frac{\sqrt{E_T}}{M}e^{-j2\pi m n/M}$, which has the additional benefit of balancing training power equally over each transmit antenna in each training symbol.  

In Appendix~\ref{section:mmseestimation}, we show that the estimation errors of each channel coefficient are equal under the assumption of i.i.d. channels with variance $\sigma_H^2$, taking the value
\begin{equation}\label{eqn:sigmae}  
\sigma_e^2=\left(\sigma_H^{-2}+\frac{1}{\sigma_n^2}\frac{E_T}{M}\right)^{-1}.
\end{equation} 
As we illustrate in Section~\ref{sub:robustprecoderdesign}, the assumption of equal estimation error variance is critical in maintaining convexity of the virtual uplink sum-MSE minimization problem.

\subsection{Linearly Precoded Data Communication Model\label{sub:lpdcm}}
Following training, we assume that all of the remaining $n_D=n-M$ symbol periods in each block will be used to broadcast data symbols.  Under the block fading assumption, the channel $\mathbf{H}$ does not change during these $n_D$ data transmissions; thus, we can design a single precoder/decoder pair to be used for all transmissions in the block.  It follows that the remaining available energy to be used for data $(E_D=E_{\max}-E_T)$ should be divided equally over the $n_D$ data transmissions, resulting in a maximum per-symbol transmit power $P_{D} = (E_{\max}-E_T)/n_D$.  

During each data transmission, user $k$ receives $L_k$ data symbols $\mathbf{x}_{k}=\left[x_{k1}, \dots, x_{kL_{k}}\right]^T$ from the base station, and the vector $\mathbf{x} = \left[\mathbf{x}_1^T, \dots, \mathbf{x}_K^T\right]^T$ comprises independent symbols with unit average energy ($\mathbb{E}\left[\mathbf{x}\mathbf{x}^{H}\right]=\mathbf{I}_{L}$, where $L=\sum_{k=1}^K L_k$).  User $k$'s data streams are precoded by the $M\times L_k$ transmit filter $\mathbf{U}_{k} = \left[\mathbf{u}_{k1}, \dots, \mathbf{u}_{kL_k}\right]$, where $\mathbf{u}_{kj}$ is the precoding beamformer for stream $j$ of user $k$ with $\|\mathbf{u}_{kj}\|_2 = 1$, and the precoders are combined in the $M\times L$ global transmitter precoder matrix $\mathbf{U} =
\left[\mathbf{U}_{1}, \dots, \mathbf{U}_{K}\right]$. Power is allocated to user $k$'s data streams in the vector $\mathbf{p}_{k}=\left[p_{k1}, \dots, p_{kL_{k}} \right]^T$ and $\mathbf{P}_k=\mathrm{diag}\left[\mathbf{p}_k\right]$; we define the downlink power allocation matrix as $\mathbf{P}=\mathrm{diag}\left[\mathbf{p}_1^T, \dots, \mathbf{p}_K^T\right]$ with $\mathrm{tr}\left[\mathbf{P}\right] \le P_D$.  Based on this model, user $k$ receives a length-$N_{k}$ vector
$\mathbf{y}_{k}^{DL}=\mathbf{H}_{k}^{H}\mathbf{U}\sqrt{\mathbf{P}}\mathbf{x} + \mathbf{n}_{k}$, where the superscript $^{DL}$ indicates the downlink, and $\mathbf{n}_k \sim \mathcal{CN}(\mathbf{0},\sigma_n^2\mathbf{I}_{N_k})$.  To estimate its $L_{k}$ symbols $\mathbf{x}_{k}$, user $k$ applies the $L_{k}\times N_{k}$ receive filter $\mathbf{V}_{k}^{H}$, yielding the estimated symbols $\hat{\mathbf{x}}_{k}^{DL}=\mathbf{V}_{k}^{H}\mathbf{H}_{k}^{H}\mathbf{U}\sqrt{\mathbf{P}}\mathbf{x}+\mathbf{V}_{k}^{H}\mathbf{n}_{k}$.

In order to design the sum-MSE minimizing precoder for the downlink, we use the virtual uplink, also illustrated in Fig.~\ref{fig:systemmodel}, where each matrix is replaced by its conjugate transpose.  We emphasize that the virtual uplink is only a mathematical construct to be used for precoder design, and that its use does not require reciprocity of the true uplink and downlink channels.  We imagine that transmissions from mobile user $k$ in the virtual uplink propagate via the \textit{flipped channel} $\mathbf{H}_k$ to the base station.  The transmit and receive filters for user $k$ become $\mathbf{V}_{k}$ and $\mathbf{U}_{k}^{H}$ respectively, with normalized precoding beamformers; i.e., $\|\mathbf{v}_{kj}\|_2 = 1$, and the uplink precoder matrices are gathered as a block diagonal matrix $\mathbf{V} = \mathrm{diag}\left[\mathbf{V}_1,\ldots,\mathbf{V}_K \right]$.  Power is allocated to user $k$'s data streams as $\mathbf{q}_{k}=\left[q_{k1}, \dots, q_{kL_{k}}\right]^T$, with $\mathbf{Q}_k=\mathrm{diag}\left[\mathbf{q}_k\right]$, $\mathbf{Q}=\mathrm{diag}\left[\mathbf{q}_1^T, \dots, \mathbf{q}_K^T\right]$, and $\mathrm{tr}\left[\mathbf{Q}\right] \le P_D$.  The received symbol vector at the base station and the estimated symbol vector for user $k$ are $\mathbf{y}^{UL} = \mathbf{HV}\sqrt{\mathbf{Q}}\mathbf{x}+\mathbf{n}=\sum_{i=1}^{K}\mathbf{H}_{i}\mathbf{V}_{i}\sqrt{\mathbf{Q}_{i}}\mathbf{x}_{i}+\mathbf{n}$ and $\hat{\mathbf{x}}_{k}^{UL} = \mathbf{U}_{k}^{H}\mathbf{y}^{UL}$, respectively, with $\mathbf{n} \sim \mathcal{CN}(\mathbf{0},\sigma_n^2\mathbf{I}_{M})$.

\subsection{Robust Convex Minimum Sum-MSE Precoder Design\label{sub:robustprecoderdesign}}
The MSE matrix for user $k$ in the virtual uplink can be written as
\begin{equation}\label{eqn:msevul}
\begin{split}
\mathbf{\varepsilon}^{UL}_k & = \mathbb{E}_{\mathbf{E},\mathbf{x},\mathbf{n}}\left[\left(\hat{\mathbf{x}}_k^{UL} -
\mathbf{x}_k\right)\left(\hat{\mathbf{x}}_k ^{UL}- \mathbf{x}_k\right)^H\right] \\
& = \mathbb{E}_{\mathbf{E}}\left[\mathbf{U}_{k}^{H}\left(\mathbf{HVQV}^H\mathbf{H}^H + \sigma_n^2\mathbf{I}\right)\mathbf{U}_{k} \right.\\
& \hspace{0.74cm} \left. -\mathbf{U}_{k}^{H}\mathbf{H}_k\mathbf{V}_k\sqrt{\mathbf{Q}_k} -\sqrt{\mathbf{Q}_k}\mathbf{V}_k^{H}\mathbf{H}_k^H\mathbf{U}_{k} + \mathbf{I}_{L_k}\right]\\
& = \mathbf{U}_{k}^{H}\tilde{\mathbf{R}}\mathbf{U}_{k}-\mathbf{U}_{k}^{H}\hat{\mathbf{H}}_k\bar{\mathbf{V}}_k - \bar{\mathbf{V}}_k^{H}\hat{\mathbf{H}}_k^H\mathbf{U}_{k} + \mathbf{I}_{L_k},
\end{split}
\end{equation}
where $\bar{\mathbf{V}}_k = \mathbf{V}_k\sqrt{\mathbf{Q}_k}$, $\tilde{\mathbf{R}}=\hat{\mathbf{H}}\mathbf{VQV}^H \hat{\mathbf{H}}^H + \sigma_{\mathrm{eff}}^2\mathbf{I}_M$.  Here, we have defined the effective noise power $\sigma_{\mathrm{eff}}^2 = \sigma_n^2 + \sum_{k=1}^K \sigma_k^2 \mathrm{tr}\left[\mathbf{V}_k\mathbf{Q}_k\mathbf{V}_k^H\right]$, under the general model with different estimation error variances $\sigma_k^2$ for each user $k$.  We have also assumed the independence of data symbols, noise, and estimation errors.  The optimum robust virtual uplink receiver for user $k$ is found using the MMSE (Wiener) filter $\tilde{\mathbf{U}}_k^H=\bar{\mathbf{V}}_k^H\hat{\mathbf{H}}_k^H\tilde{\mathbf{R}}^{-1}$.  The resulting (minimum) sum-MSE is
\begin{equation}\label{eqn:smseul}
\begin{split}
\mathrm{SMSE}_{UL} & = \sum_{k=1}^K L_k - \mathrm{tr}\left[\tilde{\mathbf{R}}^{-1}\sum_{k=1}^K \hat{\mathbf{H}}_k\bar{\mathbf{V}}_k\bar{\mathbf{V}}_k^H\hat{\mathbf{H}}_k^H\right]\\
& = L-M+\sigma_{\mathrm{eff}}^2\mathrm{tr}\left[\tilde{\mathbf{R}}^{-1}\right]
\end{split}
\end{equation}
which follows from $\mathrm{tr}\left[\mathbf{AB}\right]=\mathrm{tr}\left[\mathbf{BA}\right]$, linearity of the trace operator, and the definition of $\tilde{\mathbf{R}}$.  Since the beamforming vectors $\mathbf{v}_{kj}$ have unit norm, it follows that $\mathrm{tr}\left[\mathbf{V}_j\mathbf{Q}_j\mathbf{V}_j^H\right] =  \sum_{l=1}^{L_j} q_{jl} = \|\mathbf{q}_j\|_1$ is the sum of powers allocated to user $j$'s data streams.  Under a sum-power constraint with a maximum transmit power of $P_D$, the non-convex virtual uplink sum-MSE minimization problem can be formally defined as
\begin{equation}\label{eqn:vulOpt}
\begin{split}
\left(\mathbf{V}^*,\mathbf{Q}^*\right) = & \argmin_{\mathbf{V},\mathbf{Q}} \left(\sigma_n^2+\sum_{k=1}^K\sigma_k^2\|\mathbf{q}_k\|_1\right)\mathrm{tr}\left[\tilde{\mathbf{R}}^{-1}\right]\\
\mathrm{s.t.} \quad & q_{kl} \ge 0 \quad k=1,\dots,K; \;\; l=1,\dots,L_k,\\
& \mathrm{tr}\left[\mathbf{Q}\right] \le P_D.
\end{split}
\end{equation}
When the channel estimation error variances are equal ($\sigma_k^2 = \sigma_e^2$), the effective noise becomes $\sigma_{\mathrm{eff}}^2 = \sigma_n^2 + \sigma_e^2\sum_{k}\|\mathbf{q}_k\|_1$.  Since the minimum sum-MSE is a non-increasing function of $\sum_{k}\|\mathbf{q}_k\|_1$, we can assume that all available power allocated to data transmission will be used~\cite{DingPhD}.  Thus, the effective noise can be further simplified as $\sigma_{\mathrm{eff}}^2 = \sigma_n^2 + \sigma_e^2 P_D$ for the optimum precoder, which is no longer a function of the uplink power allocations $q_{kl}$.  The optimization problem~(\ref{eqn:vulOpt}) thus becomes convex (the minimization of $\mathrm{tr}\left[\tilde{\mathbf{R}}^{-1}\right]$ under a sum power constraint), and can thus be solved using the algorithm from~\cite{KTA06} designed for the perfect CSI case by substituting the effective noise $\sigma_{\mathrm{eff}}^2$ for the noise term $\sigma_n^2$ in the original design.

\section{Joint Optimization of Energy and Precoder Design\label{section:jointopt}}
The previous section describes the design of a robust minimum sum-MSE precoder for a fixed data power allocation, $P_D$.  In this section, we extend this result by jointly optimizing the available training and data energy with the precoder design.  As explained in Section~\ref{sub:lpdcm}, the optimum strategy for sharing the available data energy $E_D$ over $n_D$ transmitted symbols is with equal energy in each transmission.  Using this strategy, and substituting the estimation error variance from~(\ref{eqn:sigmae}) into the effective noise variance, we define the joint optimization problem
\begin{equation}\label{eqn:optDesign}
\begin{split}
& \left(\mathbf{V}^*,\mathbf{Q}^*,E_T^* \right) = \argmin_{\mathbf{V},\mathbf{Q},E_T} \sigma_{\mathrm{eff}}^2\mathrm{tr}\left[\tilde{\mathbf{R}}^{-1}\right]\\
\mathrm{s.t.} &  \qquad q_{kl} \ge 0 \quad k=1,\dots,K; \;\; l=1,\dots,L_k,\\
& \mathrm{tr}\left[\mathbf{Q}\right] = P_D, \quad P_D = \frac{E_{\max}-E_T}{n_D},\\ 
& \quad \sigma_{\mathrm{eff}}^2 = \sigma_n^2 + \frac{P_D}{\left(\sigma_H^{-2}+\frac{1}{\sigma_n^2}\frac{E_T}{M}\right)}.
\end{split}
\end{equation}

\begin{theorem}
The optimum training energy $E_T^*$ is
\begin{equation}\label{eqn:optenergy}
E_T^* = \left\{\begin{array}{cc}
\frac{E_{\max}\sqrt{M} - \frac{\sigma_n^2}{\sigma_H^2}M\sqrt{n_D}}{\sqrt{n_D} + \sqrt{M}} & E_{\max} > \frac{\sigma_n^2}{\sigma_H^2}\sqrt{M n_D}\\
&\\
0 & \mathrm{otherwise}.
\end{array}
\right.
\end{equation}
\end{theorem}
\begin{IEEEproof}
\renewcommand{\IEEEQED}{}
See Appendix~\ref{section:optTrainingEnergy}.
\end{IEEEproof}

\begin{corollary}
The optimization of training/data energy allocation and the optimum precoder design in problem~(\ref{eqn:optDesign}) are separable problems.  This result can be seen directly in~(\ref{eqn:optenergy}), as the optimum value of $E_T$ is neither a function of $\mathbf{V}$ nor $\mathbf{Q}$.
\end{corollary}
\begin{corollary}
The sum-MSE minimizing precoder can be designed using existing algorithms by setting the sum power constraint $\mathrm{tr}\left[\mathbf{Q}\right] \le P_D = \left(E_{\max}-E_T\right)/n_D$ and the noise power term to the effective noise power $\sigma_{\mathrm{eff}}^2 = \sigma_n^2 + \sigma_e^2 P_D$.
\end{corollary}
\begin{corollary}\label{corollary:threshold}
No information can be communicated using the proposed algorithm in the case where $E_{\max} \le \frac{\sigma_n^2}{\sigma_H^2}\sqrt{Mn_D}$.   If the total available energy fails to exceed this threshold, there is zero energy allocated to training; as a result, the estimated channel is $\hat{\mathbf{H}}=\mathbf{0}$ and the resulting symbol estimates are $\hat{\mathbf{x}}^{DL}=\mathbf{0}$ as well.  It is difficult to provide an intuitive understanding of this result, as we do not have a closed-form expression for the minimum sum-MSE as a function of $E_T$; however, we have observed in simulations that when $E_{\max}$ falls below the threshold, the resulting minimum sum-MSE is an increasing function of $E_T$.  It follows that the ``best'' (i.e., sum-MSE minimizing) strategy is to avoid training.

We can reinterpret this threshold result in the context of average received SNR.  If we define the average transmitted power as $P_{\mathrm{avg}}\doteq E_{\max}/n$, we can rewrite the constraint as 
\begin{equation}\label{eqn:zerocommrewrite}
SNR_{\mathrm{rx}} \doteq \frac{P_{\mathrm{avg}}\sigma_H^2}{\sigma_n^2} \le \frac{\sqrt{Mn_D}}{n_D+M}.
\end{equation}
It follows that as $n \rightarrow \infty$, a strictly positive optimum training power allocation is always feasible.  Furthermore, the largest average received SNR value that the threshold can take on is ${SNR}_{\mathrm{rx}}=-3\mathrm{dB}$, corresponding to the maximum value of the RHS of~(\ref{eqn:zerocommrewrite}) when $n_D=M$.
\end{corollary}

\section{Numerical Examples\label{section:simresults}}
We now present both analytical and simulation results to illustrate the behaviour and performance of the proposed algorithm. In these results, the flat Rayleigh fading channels are modelled with $\sigma_H^2=1$.  We scale the total energy $E_{\max}$ proportionally to the block-length $n$ to reflect a realistic average power constraint, $P_{\mathrm{avg}}=E_{\max}/n=\alpha$; in these simulations, we illustrate the case of $\alpha=1$.  As such, we define the average transmit SNR as $P_{\mathrm{avg}}/\sigma_n^2$, and find different SNR values by varying the noise power $\sigma_n^2$.  These preliminary results illustrate performance in a system with $K=2$ users, $M=4$ base station antennas, and $N_1=N_2=L_1=L_2=2$ receive antennas and data streams per user.

\begin{figure}
\hspace{-0.1in}\includegraphics[width=3.9in]{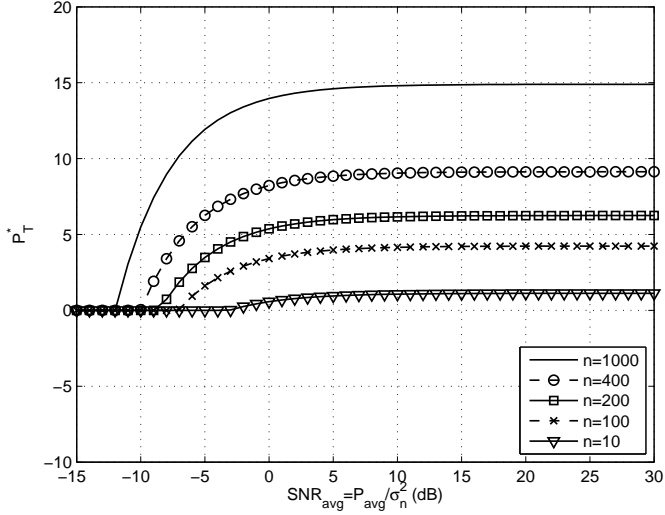}
\caption{Optimum training power $P_T^*$ for varying block length $n$}\label{fig:optimumPower}
\end{figure}

Figure~\ref{fig:optimumPower} illustrates how the optimum power allocated to training, $P_T^*$, grows with average SNR and with block length $n$.  We observe that as $n$ grows, the optimum power allocated to training becomes significantly larger than the equal power allocation $P_T=1$; however, $P_T^*$ converges fairly rapidly with increasing SNR.  We also observe the threshold behaviour described in Corollary~\ref{corollary:threshold}.
%
%

\begin{figure}
\hspace{-0.1in}\includegraphics[width=3.9in]{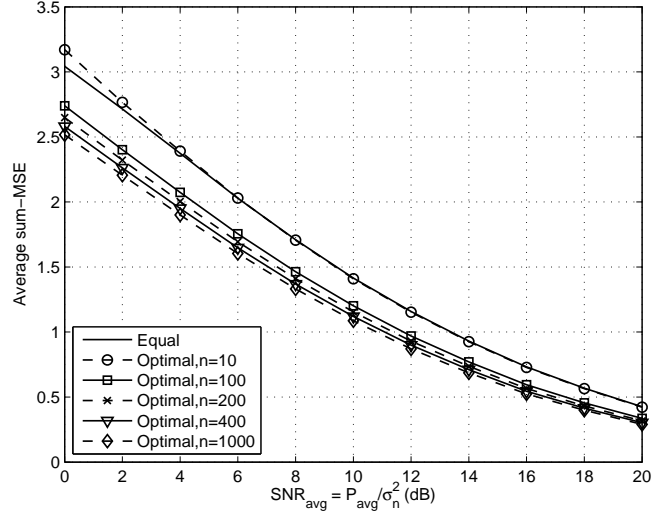}
\caption{Sum-MSE performance for equal and optimal energy allocations}\label{fig:SMSE}
\end{figure}

\begin{figure}
\hspace{-0.1in}\includegraphics[width=3.9in]{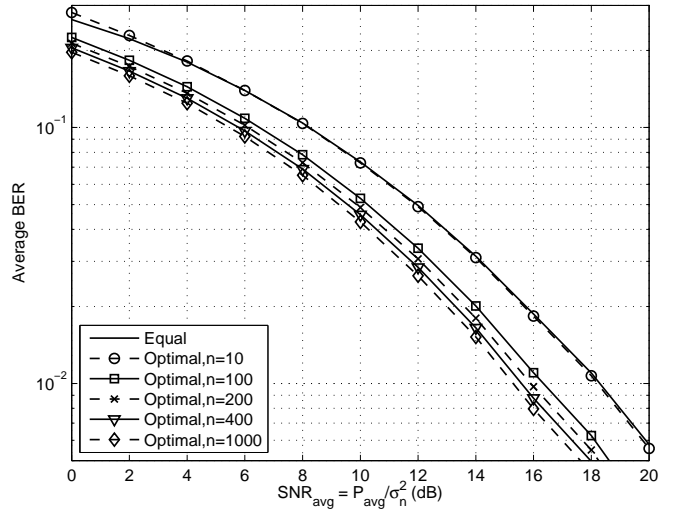}
\caption{Average BER performance for equal and optimal energy allocations}\label{fig:BER}
\end{figure}

Figures~\ref{fig:SMSE} and~\ref{fig:BER} illustrate the sum-MSE and average BER performance of the proposed algorithm.  Results in each of these plots are generated using 5000 channel realizations per average SNR value, and data symbols are
generated as uncoded QPSK.  Here, we compare performance of the proposed algorithm to the case where equal power is allocated to both training and data symbols (i.e. $P_T=P_D=1$).  We observe notable performance improvements for large block lengths ($n \gg M$), with approximately $3$ dB of SNR gain for $n=1000$.
%
%
%
%

\section{Conclusions\label{section:conclusions}}
In this paper, we have considered the problem of allocating energy to training and data symbols for systems using minimum sum-MSE linear precoding in the multiuser MIMO downlink.  We have derived the optimum closed-form energy allocation for the case of MMSE channel estimation when all users have statistically identical channels.  Furthermore, we have proven separability of the energy allocation and precoder designs; thus, existing algorithms for minimum sum-MSE precoding can be applied following energy optimization.  Preliminary simulation results demonstrate that significant improvements in performance can be made  for both realistic channel coherence intervals and transmit SNR levels.

\appendices
\section{MMSE Channel Estimation Error\label{section:mmseestimation}}
The minimum MSE matrix for the estimation of $\mathbf{h}$ can be written as
\begin{equation}\label{eqn:mseestimation}
\begin{split}
&\mathbf{\varepsilon}_{\mathrm{MMSE,est}} = \mathbb{E}_{\mathbf{h},\mathbf{n}}\left[\left(\hat{\mathbf{h}}_{\mathrm{MMSE}}-\mathbf{h}\right)\left(\hat{\mathbf{h}}_{\mathrm{MMSE}}-\mathbf{h}\right)^H\right] \\
&=\sigma_H^2\left[\mathbf{A}_0^H\left(\mathbf{X}_T^H\mathbf{X}_T + \frac{\sigma_n^2}{\sigma_H^2}\mathbf{I}\right)\mathbf{A}_0 -\left(\mathbf{A}_0^H\mathbf{X}_T^H + \mathbf{X}_T\mathbf{A}_0\right) + \mathbf{I}\right]\\
&=\sigma_H^2\left(\mathbf{I}-\mathbf{X}_T^H\left(\mathbf{X}_T^H\mathbf{X}_T + \frac{\sigma_n^2}{\sigma_H^2}\mathbf{I}_{n_T}\right)^{-1}\mathbf{X}_T\right)\\
&=\sigma_H^2\left(\mathbf{I}+\frac{\sigma_H^2}{\sigma_N^2}\mathbf{X}_T\mathbf{X}_T^H\right)^{-1}\\
&=\left(\sigma_H^{-2}+\frac{1}{\sigma_n^2}\frac{E_T}{M}\right)^{-1}\mathbf{I},
\end{split}
\end{equation}
where we have assumed that $\mathbf{h}$ and $\mathbf{z}$ are independent.  The fourth equality follows from application of the matrix inversion lemma, $\left(\mathbf{A}+\mathbf{BCD}\right)^{-1}=\mathbf{A}^{-1}-\mathbf{A}^{-1}\mathbf{B}\left(\mathbf{C}^{-1}+\mathbf{DA}^{-1}\mathbf{B}\right)^{-1}\mathbf{DA}^{-1}$.  Since the estimation error $\hat{\mathbf{h}}_{\mathrm{MMSE}}-\mathbf{h}$ is a linear combination of random vectors from a multivariate Gaussian distribution with uncorrelated components, it follows that the estimation errors are also independent Gaussian random variables.

\section{Optimum Training and Data Energy Allocation\label{section:optTrainingEnergy}}
Here, we derive a closed-form expression for the optimum training energy $E_T^*$ that minimizes the sum-MSE precoder design under a sum-energy constraint, $E_T+E_D \le E_{\max}$.  Due to space limitations, we are only able to show the most common case of long blocks (with $n \gg M$, and consequently $n_D > M$); however, the identical result applies for $n_D \le M$.

We perform the optimization in terms of the training power $P_T=E_T/M$.  Using the virtual uplink MSE from~(\ref{eqn:smseul}) as the objective function, and the energy constraints $E_T \ge 0$ and $E_T \le E_{\max}$, we derive the Karush-Kuhn-Tucker (KKT) conditions
\begin{align}
& \frac{\partial \mathrm{SMSE}_{UL}}{\partial P_T} + \lambda_{\max}M - \lambda_{+} = 0 \hspace{-3cm}&\label{eqn:KKTstationarity}\\
& P_TM \ge 0, & P_TM \le E_{\max}\label{eqn:KKTprimal}\\
& \lambda_{+} \ge 0, & \lambda_{\max} \ge 0\label{eqn:KKTdual}\\
& \lambda_{+}P_TM = 0, & \lambda_{\max}\left(P_TM-E_{\max}\right)=0.\label{eqn:KKTcs}
\end{align}
We consider only the solutions where the constraints are not binding, as allowing either constraint to hold with equality prevents us from reaching a global minimum for the optimization problem.  When $P_TM=0$, no training symbols are sent, and the resulting channel estimate is $\hat{\mathbf{H}}^H=\mathbf{0}$.  If $P_TM=E_{\max}$, zero energy remains for data transmission.  In either of these cases, the resulting data symbol estimates are $\hat{\mathbf{x}}_k^{UL}=\mathbf{0}$, and no information can be communicated.  Since neither constraint is binding, complementary slackness~(\ref{eqn:KKTcs}) requires that $\lambda_{\max}=\lambda_{+}=0$; thus, any minimizer can be found by considering the unconstrained minimization of $\mathrm{SMSE}_{UL}$ and checking feasibility of the resulting solutions.  We begin by rewriting the effective noise power,
\begin{equation}\label{eqn:sigmaeff}
\begin{split}
\sigma_{\mathrm{eff}}^2 & = \sigma_n^2 + \frac{\sigma_n^2}{n_D}\left(\frac{E_{\max}-P_TM}{\rho + P_T}\right),
\end{split}
\end{equation}
with $\rho = \sigma_n^2/\sigma_H^2$.  Define the derivative
\begin{equation}\label{eqn:dsigma}
\begin{split}
D_\sigma \doteq \frac{\partial \sigma_{\mathrm{eff}}^2}{\partial P_T} = \frac{-\sigma_n^2\left(E_{\max} + \rho M\right)}{\left(\rho + P_T\right)^2}.
\end{split}
\end{equation}
We then separate the data power $P_D$ from the uplink power allocation by rewriting $\mathbf{Q}=P_D\tilde{\mathbf{Q}}$, with associated sum power constraint $\mathrm{tr}\left[\tilde{\mathbf{Q}}\right] \le 1$.  It follows that 
\begin{equation}
\tilde{\mathbf{R}}=\left(\frac{E_{\max}-P_TM}{n_D}\right)\mathbf{HV}\tilde{\mathbf{Q}}\mathbf{V}^H\mathbf{H}^H + \sigma_{\mathrm{eff}}^2.
\end{equation}
Define the derivative of the trace function
\begin{equation}
\begin{split}
& D_{\mathrm{tr}} \doteq \frac{\partial \mathrm{tr}\left[\tilde{\mathbf{R}}^{-1}\right]}{\partial P_T} = -\mathrm{tr}\left[\tilde{\mathbf{R}}^{-1}\frac{\partial \tilde{\mathbf{R}}}{\partial P_T}\tilde{\mathbf{R}}^{-1}\right]\\
&= \mathrm{tr}\left[\tilde{\mathbf{R}}^{-2}\left(\frac{M}{n_D}\mathbf{HV}\tilde{\mathbf{Q}}\mathbf{V}^H\mathbf{H}^H - D_\sigma\mathbf{I}\right)\right]\\
&= -\mathrm{tr}\left[\tilde{\mathbf{R}}^{-2}\right]\left(D_\sigma + \frac{M\sigma_{\mathrm{eff}}^2}{n_DP_D}\right) + \frac{M}{n_DP_D}\mathrm{tr}\left[\tilde{\mathbf{R}}^{-1}\right].
\end{split}
\end{equation}

The candidate values of $P_T$ for unconstrained global optimality satisfy
\begin{equation}\label{eqn:statpoints}
\begin{split}
& \frac{\partial \mathrm{SMSE}_{UL}}{\partial P_T} = D_\sigma\mathrm{tr}\left[\tilde{\mathbf{R}}^{-1}\right] + \sigma_{\mathrm{eff}}^2 D_{\mathrm{tr}} = 0\\
&= \left(\mathrm{tr}\left[\tilde{\mathbf{R}}^{-1}\right]  - \sigma_{\mathrm{eff}}^2\mathrm{tr}\left[\tilde{\mathbf{R}}^{-2}\right] \right)\left(D_\sigma + \frac{M\sigma_{\mathrm{eff}}^2}{n_DP_D}\right).
\end{split}
\end{equation}
The first term in~(\ref{eqn:statpoints}) can be rewritten as $P_D\mathrm{tr}\left[\tilde{\mathbf{R}}^{-1}\mathbf{HV}\tilde{\mathbf{Q}}\mathbf{V}^H\mathbf{H}^H\tilde{\mathbf{R}}^{-1}\right]$, which only has a trivial zero $P_T=E_{\max}/M$ (corresponding to $P_D=0$) since the argument of the trace function is positive definite for non-zero power allocations $\mathbf{Q}$.  Any globally optimum $P_T^*$ must therefore satisfy
\begin{equation}\label{eqn:zeroterm}
D_\sigma + \frac{M\sigma_{\mathrm{eff}}^2}{n_DP_D}=0.
\end{equation}
Substituting the definitions of~(\ref{eqn:sigmaeff}) and~(\ref{eqn:dsigma}) gives rise to the following quadratic equation in $P_T$,
\begin{equation}\label{eqn:quadratic}
P_T^2\left(n_D - M\right) + 2 P_T\left(E_{\max} + \rho n_D\right) = \frac{E_{\max}^2}{M} - \rho^2n_D.
\end{equation}
The two roots of this quadratic equation are 
\begin{equation}
P_T = \frac{1}{n_D-M}\left(-E_{\max} - \rho n_D \pm \gamma\right),
\end{equation}
with
\begin{equation}
\begin{split}
\gamma & \doteq \sqrt{n_D\left(\rho^2M + 2\rho E_{\max} + \frac{E_{\max}^2}{M}\right)}\\
& = E_{\max}\sqrt{\frac{n_D}{M}} + \rho\sqrt{n_DM}
\end{split}
\end{equation}

Clearly, for $n_D > M$, the negative root ($-\gamma$) results in an infeasible solution $P_T < 0$.  We can see that the positive root gives rise to
\begin{equation}
\begin{split}
P_T^* & = \frac{E_{\max}\left(\sqrt{\frac{n_D}{M}} - 1\right) - \rho n_D\left(1-\sqrt{\frac{M}{n_D}}\right)}{n_D-M}\\
&= \frac{E_{\max}\left(\frac{\sqrt{n_D}-\sqrt{M}}{\sqrt{M}}\right) - \rho n_D\left(\frac{\sqrt{n_D}-\sqrt{M}}{\sqrt{n_D}}\right)}{\left(\sqrt{n_D}-\sqrt{M}\right)\left(\sqrt{n_D}+\sqrt{M}\right)}\\
&= \frac{\frac{E_{\max}}{\sqrt{M}} - \rho\sqrt{n_D}}{\sqrt{n_D}+\sqrt{M}}.
\end{split}
\end{equation}
This solution always satisfies $P_T^* M < E_{\max}$, and is only infeasible (with $P_T^* < 0$) if $E_{\max} < \rho\sqrt{n_DM}$.  

Finally, we prove that this stationary point $P_T^*$ is indeed a global minimum.  We observe that the second derivative of $\mathrm{SMSE}_{UL}$ can be written as
\begin{equation}\label{eqn:d2smse}
\begin{split}
& \mathrm{tr}\left[\tilde{\mathbf{R}}^{-1}\mathbf{HVQ}\mathbf{V}^H\mathbf{H}^H\tilde{\mathbf{R}}^{-1}\right]\frac{\partial}{\partial P_T}\left(D_\sigma + \frac{M\sigma_{\mathrm{eff}}^2}{n_DP_D}\right)\\
& + \left(D_\sigma + \frac{M\sigma_{\mathrm{eff}}^2}{n_DP_D}\right)\frac{\partial}{\partial P_T}\left(\mathrm{tr}\left[\tilde{\mathbf{R}}^{-1}\mathbf{HVQ}\mathbf{V}^H\mathbf{H}^H\tilde{\mathbf{R}}^{-1}\right]\right),
\end{split}
\end{equation}
but the second term vanishes at $P_T^*$ due to~(\ref{eqn:zeroterm}).  We previously showed that the trace term is strictly positive; thus, to prove that $P_T^*$ is a global minimizer, we must only show that the remaining term in the second derivative is positive at $P_T^*$:
\begin{equation}
\begin{split}
\frac{\partial}{\partial P_T}\left(D_\sigma + \frac{M\sigma_{\mathrm{eff}}^2}{n_DP_D}\right) & = \frac{\partial D_\sigma}{\partial P_T} + \frac{M D_\sigma}{n_DP_D} + \frac{M^2\sigma_{\mathrm{eff}}^2}{n_D^2P_D^2}\\
& = \frac{\partial D_\sigma}{\partial P_T} + \frac{M}{n_DP_D}\left(D_\sigma + \frac{M\sigma_{\mathrm{eff}}^2}{n_DP_D}\right).
\end{split}
\end{equation}
At the point $P_T=P_T^*$, the second term vanishes due to~(\ref{eqn:zeroterm}).  The remaining term 
\begin{equation}
\left. \frac{\partial D_\sigma}{\partial P_T}\right|_{P_T=P_T^*} =  \frac{2\sigma_n^2\left(E_{\max} + \rho M\right)}{\left(\rho + P_T^*\right)^3},
\end{equation}
is positive; thus, the training power $P_T^*$ is the global minimizer.
\QED

\bibliographystyle{IEEEtran}
\bibliography{IEEEabrv,AdamCISS}

\begin{thebibliography}{10}
\providecommand{\url}[1]{#1}
\csname url@samestyle\endcsname
\providecommand{\newblock}{\relax}
\providecommand{\bibinfo}[2]{#2}
\providecommand{\BIBentrySTDinterwordspacing}{\spaceskip=0pt\relax}
\providecommand{\BIBentryALTinterwordstretchfactor}{4}
\providecommand{\BIBentryALTinterwordspacing}{\spaceskip=\fontdimen2\font plus
\BIBentryALTinterwordstretchfactor\fontdimen3\font minus
  \fontdimen4\font\relax}
\providecommand{\BIBforeignlanguage}[2]{{%
\expandafter\ifx\csname l@#1\endcsname\relax
\typeout{** WARNING: IEEEtran.bst: No hyphenation pattern has been}%
\typeout{** loaded for the language `#1'. Using the pattern for}%
\typeout{** the default language instead.}%
\else
\language=\csname l@#1\endcsname
\fi
#2}}
\providecommand{\BIBdecl}{\relax}
\BIBdecl

\bibitem{SSJB05}
M.~Schubert, S.~Shi, E.~A. Jorswieck, and H.~Boche, ``Downlink sum-{MSE}
  transceiver optimization for linear multi-user {MIMO} systems,'' in
  \emph{Proc. Asilomar Conf. on Signals, Systems and Computers}, Monterey, CA,
  Sep. 2005, pp. 1424--1428.

\bibitem{KTA06}
A.~M. Khachan, A.~J. Tenenbaum, and R.~S. Adve, ``Linear processing for the
  downlink in multiuser {MIMO} systems with multiple data streams,'' in
  \emph{Proc. {IEEE} Internat. Conf. on Communications (ICC 06)}, Istanbul,
  Turkey, Jun. 2006.

\bibitem{MJHU06}
A.~Mezghani, M.~Joham, R.~Hunger, and W.~Utschick, ``Transceiver design for
  multi-user {MIMO} systems,'' in \emph{Proc. {ITG/IEEE} Workshop on Smart
  Antennas}, Ulm, Germany, Mar. 2006.

\bibitem{SSB07}
S.~Shi, M.~Schubert, and H.~Boche, ``Downlink {MMSE} transceiver optimization
  for multiuser {MIMO} systems: Duality and sum-{MSE} minimization,''
  \emph{{IEEE} Trans. Signal Process.}, vol.~55, no.~11, pp. 5436--5446, Nov.
  2007.

\bibitem{DingPhD}
M.~Ding, ``Multiple-input multiple-output wireless system designs with
  imperfect channel knowledge,'' Ph.D. dissertation, Queen's University, Jul.
  2008.

\bibitem{SD08}
M.~B. Shenouda and T.~N. Davidson, ``On the design of linear transceivers for
  multiuser systems with channel uncertainty,'' \emph{{IEEE} J. Sel. Areas
  Commun.}, vol.~26, no.~6, pp. 1015--1024, Aug. 2008.

\bibitem{DB09}
M.~Ding and S.~D. Blostein, ``{MIMO} minimum total {MSE} transceiver design
  with imperfect {CSI} at both ends,'' \emph{{IEEE} Trans. Signal Process.},
  vol.~57, no.~3, pp. 1141--1150, Mar. 2009.

\bibitem{80216e}
\emph{{IEEE} 802.16e Standard, Part 16: Air Interface for Fixed Broadband
  Wireless Access Systems}.\hskip 1em plus 0.5em minus 0.4em\relax New York,
  NY, USA: {IEEE}, 2005.

\bibitem{H08}
J.~C. Haartsen, ``Impact of non-reciprocal channel conditions in broadband
  {TDD} systems,'' in \emph{Proc. {IEEE} {PIMRC} '08}, Cannes, France, Sep.
  2008.

\bibitem{TAY08}
A.~J. Tenenbaum, R.~S. Adve, and Y.-S. Yuk, ``Channel prediction and feedback
  in multiuser broadcast channels,'' in \emph{Proc. {IEEE} {CWIT} '09}, Ottawa,
  ON, May 2008, pp. 67--70.

\bibitem{BG06}
M.~Biguesh and A.~B. Gershman, ``Training-based {MIMO} channel estimation: A
  study of estimator tradeoffs and optimal training signals,'' \emph{{IEEE}
  Trans. Signal Process.}, vol.~54, no.~3, pp. 884--893, Mar. 2006.

\end{thebibliography}
\end{document}